\documentclass[]{pasj01}

\Received{}
\Accepted{}
 
\usepackage{natbib}
\usepackage{url}
\newcommand{\HI}{\mbox{H\,{\sc i}}}
\newcommand{\oii}{\mbox{[O\,{\sc ii}]}}
\newcommand{\nii}{\mbox{[N\,{\sc ii}]}}
\newcommand{\oiii}{\mbox{[O\,{\sc iii}]}}
\newcommand{\caii}{\mbox{Ca\,{\sc ii}}}

\begin{document} 

\title{ 
Complex distribution and velocity field of molecular gas in NGC~1316 as revealed by Morita Array of ALMA
}

\author{Kana \textsc{Morokuma-Matsui}\altaffilmark{1,6}}%
\altaffiltext{1}{Institute of Space and Astronautical Science/Japan Aerospace Exploration Agency, 3-1-1 Yoshinodai, Chuo-ku, Sagamihara, Kanagawa 252-5210, Japan}
\email{kanamoro@ioa.s.u-tokyo.ac.jp}

\author{Paolo \textsc{Serra}\altaffilmark{2}}
\author{Filippo M. \textsc{Maccagni}\altaffilmark{2}}
\altaffiltext{2}{INAF -- Osservatorio Astronomico di Cagliari, Via della Scienza 5, I-09047 Selargius (CA), Italy}

\author{Bi-Qing \textsc{For}\altaffilmark{3,5}}
\altaffiltext{3}{ARC Centre of Excellence for All Sky Astrophysics in 3 Dimensions (ASTRO 3D)}

\author{Jing \textsc{Wang}\altaffilmark{4}}
\altaffiltext{4}{Kavli Institute for Astronomy and Astrophysics, Peking University, Beijing 100871, China}

\author{Kenji \textsc{Bekki}\altaffilmark{5}}
\altaffiltext{5}{ICRAR M468 The University of Western Australia 35 Stirling Hwy, Crawley Western Australia 6009, Australia}

\author{Tomoki \textsc{Morokuma}\altaffilmark{6}}
\author{Fumi \textsc{Egusa}\altaffilmark{6}}
\altaffiltext{6}{Institute of Astronomy, Graduate School of Science, The University of Tokyo, 2-21-1 Osawa, Mitaka, Tokyo 181-0015, Japan}

\author{Daniel \textsc{Espada}\altaffilmark{7,8}}
\author{Rie \textsc{Miura, E.}\altaffilmark{7}}
\author{Kouichiro \textsc{Nakanishi}\altaffilmark{7,8}}
\altaffiltext{7}{National Astronomical Observatory of Japan, National Institutes of Natural Sciences, 2-21-1 Osawa, Mitaka, Tokyo 181-8588, Japan}
\altaffiltext{8}{SOKENDAI (The Graduate University for Advanced Studies), 2-21-1 Osawa, Mitaka, Tokyo 181-8588, Japan}

\author{B\"arbel S. \textsc{Koribalski}\altaffilmark{9}}
\altaffiltext{9}{Australia Telescope National Facility, CSIRO Astronomy \& Space Science, PO Box 76, Epping, NSW 1710, Australia}

\author{Tsutomu T. \textsc{Takeuchi}\altaffilmark{10}}
\altaffiltext{10}{Division of Particle and Astrophysical Science, Nagoya University, Furo-cho, Chikusa-ku, Nagoya 464-8602, Japan}



\maketitle

\begin{abstract}
We present the results of $^{12}$CO($J$=1-0) mosaicing observations of the cD galaxy NGC~1316 at kpc-resolution performed with the Morita Array of the Atacama Large Millimeter/submillimeter Array (ALMA).
We reveal the detailed distribution of the molecular gas in the central region for the first time: a shell structure in the northwest, a barely resolved blob in the southeast of the center and some clumps between them.
The total molecular gas mass obtained with a standard Milky-Way CO-to-H$_2$ conversion factor is $(5.62 \pm 0.53)\times10^8$~M$_\odot$, which is consistent with previous studies.
The disturbed velocity field of the molecular gas suggests that the molecular gas is injected very recently ($<1$~Gyr) if it has an external origin and is in the process of settling into a rotating disk.
Assuming that a low-mass gas-rich galaxy has accreted, the gas-to-dust ratio and H$_2$-to-\HI~ratio are unusually low ($\sim 28$) and high ($\sim 5.6$), respectively.
To explain these ratios, additional processes should be taken into accounts such as an effective dust formation and conversion from atomic to molecular gas during the interaction.
We also discuss the interaction between the nuclear jet and the molecular gas.

\end{abstract}

\section{Introduction}

Galaxy major/minor merger and Active Galactic Nucleus (AGN) play an important role in galaxy evolution, especially for massive early-type galaxies \citep[ETGs, e.g.,][]{Croton:2006rt,Guo:2011aa,Oogi:2013aa,Oogi:2016aa,Rodriguez-Gomez:2016aa}.
Mergers are considered to drive an effective gas inflow into the central region of galaxies and may activate star formation and/or AGN there \citep[e.g.,][]{Hopkins:2008aa}.
Once the AGN is ignited, its glaring radiation and powerful jet are believed to suppress star formation by heating and/or blowing up the cold interstellar medium \citep[ISM, e.g.,][]{Croton:2006rt,Fabian:2012aa}.
Therefore, it is important to investigate cold ISM properties of merging and/or post-merger galaxies with AGN.

NGC~1316 is one of the best targets to study the galaxy merger and the interaction between the AGN jet and ISM.
This is a cD galaxy\footnote{
NGC~1316 is classified as a lenticular galaxy in RC3 \citep{de-Vaucouleurs:1991pr}, but it is confirmed that this galaxy is a typical D-type galaxy, and probably a cD galaxy based on deep and wide optical observation \citep{Schweizer:1980aa,Iodice:2017aa}.
}
located on the outskirt of the Fornax cluster in the south-west direction \citep[$\sim 4$ degree from the center or $\sim 2$ times virial radius,][]{Drinkwater:2001aa}.
This galaxy, also known as Fornax~A, is the third-nearest radio-bright galaxy after NGC~5128 (Centaurus~A) and M~87.
It has prominent radio lobes spanning 33 arcmin at the position angle (P.A.) of 110 degree \citep{Ekers:1983aa,McKinley:2015aa}.
The unusually low X-ray luminosity of the AGN of NGC~1316 suggests a declined activity in the last 0.1~Gyr \citep{Iyomoto:1998aa,Kim:2003aa,Lanz:2010aa}.
In the central region, there is an S-shaped nuclear radio jet, which is considered to be bent by interaction with ISM \citep{Geldzahler:1984aa}.
Observations in X-ray and vibration-rotation H$_2$ line suggest an interaction between AGN jet and ISM of NGC~1316 \citep{Kim:1998aa,Roussel:2007aa}.
A peculiar PAH spectrum of this galaxy (an unusually high intensity ratio of the 11.3~$\mu$m to the 7.7~$\mu$m emission lines and an absence of the 3.3~$\mu$m emission) is observed and considered to be due to an effective destruction of smaller PAHs through sputtering in the hot plasma \citep{Iyomoto:1998aa,Smith:2007aa,Kaneda:2007aa}.

There is much observational evidence of a rich history of merging events for NGC~1316.
Contrary to the central cD galaxy of the Fornax cluster, NGC~1399, NGC~1316 is surrounded by many late-type galaxies and considered to be in an early phase of cD-galaxy formation \citep{Iodice:2017aa}.
Deep optical observations revealed complex structures such as ripples and loops in the outer regions, as well as prominent dust extinction in the central region of the galaxy (see Figure~\ref{fig:comparison}a):
a shell structure in the northwest, a blob in the south-east, and several dust patches between them \citep[][]{Schweizer:1980aa,Schweizer:1981aa,Grillmair:1999aa,Carlqvist:2010aa,Iodice:2017aa}.
The regions with high dust extinction are bright in far-infrared (FIR) wavelengths \citep{Lanz:2010aa,Galametz:2012aa,Galametz:2014aa,Duah-Asabere:2016aa}.
\cite{Schweizer:1980aa} suggested the presence of the ``polar'' rapidly rotating ionized disk based on optical spectroscopic observations, although the author presented only a picture of 2D spectrum but not a rotation curve.
The author also claimed that the central dust patches are associated with this ionized gas disk.

Emission lines associated with rotational transition of molecules are powerful tools to investigate the velocity field of the dust patches which are traced with continuum emission, since dust and molecular gas are generally well-mixed.
In NGC~1316, CO emission is detected in the vicinity of the dust shell in the northwest and the dust blob in the southeast with single-dish observations \citep{Sage:1993aa,Horellou:2001aa}, whereas atomic gas has not been detected down to $10^8$~M$_\odot$ \citep{Horellou:2001aa}.
Total molecular gas mass is reported to be $(4.0-5.4) \times 10^8$~M$_\odot$ \citep[corrected for the choice of distance and CO-to-H$_2$ conversion factor to match that of this study,][]{Wiklind:1989aa,Sage:1993aa,Horellou:2001aa}.
The single-dish CO velocity varies by $\sim 460$ km~s$^{-1}$ when moving from 72'' northwest to 35'' southeast of the center.
This variation is smaller than that measured for the ionized gas by \cite{Schweizer:1980aa}, which is $\sim 700$ km~s$^{-1}$ when moving from 54'' northwest to 35'' southeast.
However, the complex dust structure has not been sufficiently resolved with these single-dish observations.

In this study, we present new results of $^{12}$CO($J$=1-0) mapping observations of NGC~1316 with the Morita Array (Compact array) of the Atacama Large Millimeter/submillimeter Array (ALMA) as part of ALMA $^{12}$CO($J$=1-0) survey toward 64 Fornax galaxies (project code: 2017.1.00129.S, PI: Kana Morokuma-Matsui).
We also conducted optical spectroscopic observations with the Low-Resolution Imaging Spectrometer (LRIS) at the Keck Observatory to compare the kinematics of molecular gas and ionized gas, since the previous optical study presented only a picture of 2-D spectrum.
The content of this paper is as follows:
ALMA and Keck observations and data reductions are described in section~\ref{sec:observation_reduction},
the spatial distribution of molecular gas and velocity structures of both molecular and ionized gas are presented in section~\ref{sec:results}.
We discuss the kinematic nature of the molecular gas and the possible cause for the nuclear jet bending are discussed in section~\ref{sec:discussions} and section~\ref{sec:bending}, respectively,
and summarize this study in section~\ref{sec:summary}.
We adopt the distance to NGC~1316 of $20.8\pm0.5$~Mpc \citep{Cantiello:2013aa} throughout the paper and other basic parameters of NGC~1316 are summarized in Table~\ref{tab:first}.
At this distance, 1$''$ corresponds to $\sim 100$~pc.

\begin{table}
  \tbl{Basic information of NGC~1316}{%
  \begin{tabular}{lc}
      \hline
      Coordinate & (03$^h$22$^m$41.718$^s$, -37$^\circ$12$'$29$''$.62)\footnotemark[$*$]  \\ 
      Velocity & $1732\pm10$~km~s$^{-1}$\footnotemark[$**$]  \\ 
      Redshift & $0.00587$\footnotemark[$***$]  \\ 
      Distance & $20.8\pm0.5$ Mpc\footnotemark[$\dag$]\\
      & ($1''\sim100$~pc)\\
      Stellar mass & $5.9\times10^{11}$ M$_\odot$\footnotemark[$\ddag$]\\
      SFR & $0.30-0.77$ M$_\odot$ yr$^{-1}$\footnotemark[$\ddag$]\\
      Dust mass & $2.0\times10^{7}$ M$_\odot$\footnotemark[$\sharp$]\\
      Morph. & cD\footnotemark[$\S$]\\
      \hline
    \end{tabular}}\label{tab:first}
\begin{tabnote}
\footnotemark[$*$] \cite{Shaya:1996aa};
\footnotemark[$**$] \cite{Longhetti:1998aa}, with \oii~emission, kinematic-LSR systemic velocity in radio definition;
\footnotemark[$***$] NED;
\footnotemark[$\dag$] \cite{Cantiello:2013aa};
\footnotemark[$\ddag$] \cite{Duah-Asabere:2016aa}, assuming Chabrier IMF \citep{Chabrier:2003oe} for stellar mass and Kroupa IMF \citep{Kroupa:2001fj} for SFR;
\footnotemark[$\S$] \cite{Schweizer:1980aa};
\footnotemark[$\sharp$] \cite{Draine:2007aa,Lanz:2010aa}.
\end{tabnote}
\end{table}

\section{Observations and data analysis}
\label{sec:observation_reduction}

\begin{table*}
  \tbl{Summary of the observations}{%
  \begin{tabular}{lcccc}
      \hline
      Telescope/instruments & Project & Obs. date (UT) & Angular resolution & Velocity resolution\\ 
      \hline
      ALMA/Band~3 & 2017.1.00129.S & Oct.~16~2017-May~31~2018 & $15''.5\times7''.7$ & 9.9~km~s$^{-1}$\\ 
      Keck/LRIS & PI of S.~Perlmutter (U085) & Sept.~16~2018 & $1.0''^{*}$ & $300-500$~km~s$^{-1**}$\\ 
      \hline
    \end{tabular}}\label{tab:observation}
\begin{tabnote}
\footnotemark[$*$] Slit width.\\
\footnotemark[$**$] The accuracy of determining the central velocity of the spectra is $\lesssim100$~km~s$^{-1}$.
\end{tabnote}
\end{table*}

We describe our ALMA observations and data analysis in this section, focusing on NGC~1316.
Detailed information of the whole 64 galaxy samples will be presented in the forthcoming overview paper for the survey (Morokuma-Matsui et al. in prep.).
Optical spectroscopic observations with Keck/LRIS are also described in this section.
Both the radio and optical observations are summarized in Table~\ref{tab:observation}.

\subsection{ALMA 7-m and total-power array observations}

The 7-m antennae (7M) and 12-m total power (TP) array observations in Band~3 were carried out during cycle 5 as part of the ALMA survey toward 64 Fornax galaxies.

For the 7M observations, from nine to eleven antennae were employed with baselines ranging from 8.9~m to 48.9~m (uv range from 2.5~k$\lambda$ to 12.0~k$\lambda$, angular scale\footnote{\url{https://almascience.org/about-alma/alma-basics}} from $13''$ to $74''$) during the observing campaign from October~16 to December~21 in 2017.
The number of pointings for the NGC~1316 mosaic is ten.
The 64 galaxies were divided into two scheduling blocks (SBs) and NGC~1316 is included in the SB named as ${\tt NGC1316\_a\_03\_7M}$.
Each SB consists of 33 execution blocks (EBs).
The objects J0334-4008 and J0522-3627 were observed as a phase calibrator and a bandpass/flux calibrator, respectively.
In order to cover the whole sample of Fornax galaxies with a single correlator setup, the systemic velocity for each source was fixed to 1500 km s$^{-1}$.
The $^{12}$CO($J$=1-0) emission line (rest frequency $\nu_{\rm rest}$ of 115.271202~GHz) was covered in one of the upper-side band (USB) spectral windows (SPWs) whose bandwidth and resolution are 1875~MHz ($4902.5$~km~s$^{-1}$) and 1.128~MHz ($2.952$~km~s$^{-1}$), respectively.
The other USB SPW (centered on 113.3962~GHz) and two lower-side band SPWs (centered on 103.2712~GHz and 101.3962~GHz) were mainly used for continuum observations.

The TP on-the-fly observations were carried out from May~29 to 31 in 2018 with three or four antennae.
The TP mapping area covers the whole area observed with the 7M array.
The SB name of the TP observation of NGC~1316 is ${\tt NGC1316\_a\_03\_TP}$, which consists of five EBs.
The correlator and SPW setups are the same as that of 7M observations.


\subsection{ALMA data analysis with CASA}

\begin{figure}[t]
\begin{center}
\includegraphics[width=80mm, bb=0 0 405 305]{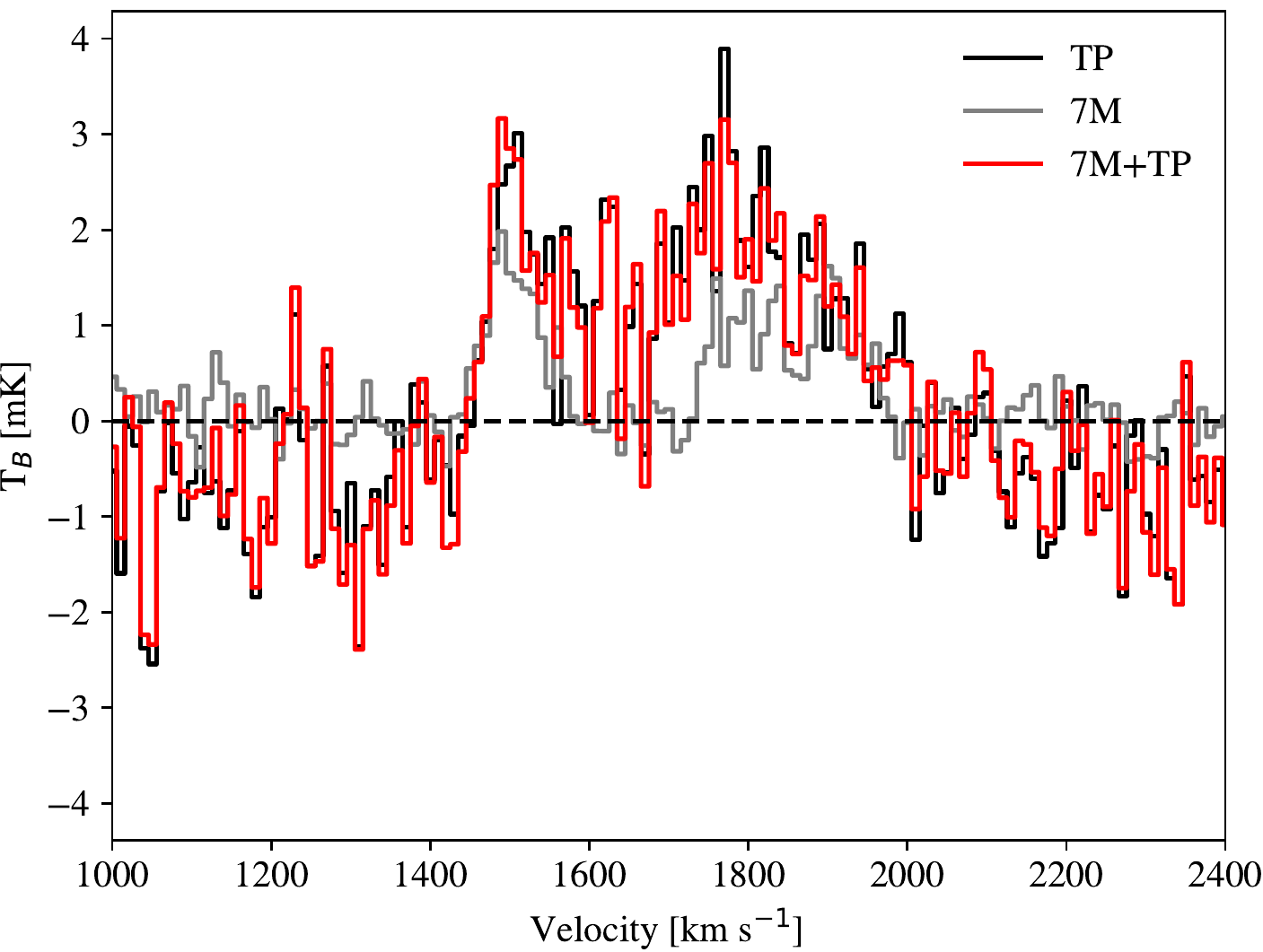}
\end{center}
\caption{Comparison of total integrated spectra of TP (black), 7M (grey) and 7M+TP data (red).
The CO spectra are generated in the box area centered on the coordinate of (R.A., Dec.)$_{\rm ICRS}=(03^{h}22^{m}41^{s}.447, -37^{\circ}12'16''.184)$ with a width of 120~arcsec and a height of 160~arcsec.
Data combination of 7M+TP successfully recovered the total flux measured with TP array.
}
\label{fig:total_spectra}
\end{figure}

Data calibration and imaging were conducted with the standard ALMA data analysis package, the Common Astronomy Software Applications \citep[CASA,][]{McMullin:2007aa,Petry:2012aa}.
The absolute flux and gain fluctuations of the 7M data were calibrated with the ALMA Science Pipeline (version of r40896 of Pipeline-CASA51-P2-B) in the CASA 5.1.1 package.
The flux accuracy of the ALMA 7M band-3 data is reported to be better than 5~\% (ALMA proposer's guide).
The obtained fluxes of the phase calibrator at individual SPWs are consistent within the errors with the values estimated using the other ALMA measurements of the same source on the closest date to our observations (e.g., our measurement of 0.46~Jy on Oct.~16 2017 vs the other measurements of $0.45\pm0.03$~Jy which is estimated based on the flux measurements at 91.46~GHz, 103.5~GHz, and 343.48~GHz on Oct.~11 2017)\footnote{We utilized {\tt getALMAFluxForMS} of the ``Analysis Utilities'' (\url{https://casaguides.nrao.edu/index.php/Analysis_Utilities}).}.
The 7M $^{12}$CO($J$=1-0) mosaic data cube was generated with {\tt TCLEAN} task in CASA version 5.4 with options of {\tt Briggs} weighting with a {\tt robust} parameter of 0.5, {\tt auto-multithresh} mask with standard values for 7M data provided in the CASA Guides for auto-masking\footnote{\url{https://casaguides.nrao.edu/index.php/Automasking_Guide}} ({\tt sidelobethreshold} of 1.25, {\tt noisethreshold} of 5.0, {\tt minbeamfrac} of 0.1, {\tt lownoisethreshold} of 2.0, and {\tt negativethreshold} of 0.0), and {\tt niter} of 10000.
The 7M and TP data are combined basically on the image plane with {\tt FEATHER} task in CASA to account for zero spacing information.
The achieved synthesized beam was $15''.2 \times 7''.7$ ($1.5~{\rm kpc} \times 0.8~{\rm kpc}$ at the distance of NGC~1316) with a P.A. of 89$^\circ$.
The beam size is a few times smaller than the $43''$ beam of the 15 m Swedish-ESO Submillimeter Telescope used in the previous CO observations of NGC~1316 \citep{Wiklind:1989aa,Sage:1993aa,Horellou:2001aa}.
After the combination, we verified that the 7M+TP data matched the spectral profile of the TP data (Figure~\ref{fig:total_spectra}).
The achieved rms noise of the box area centered on the galaxy center with a width of 80~arcsec and with a height of 105~arcsec with the final velocity resolution of 9.9~km~s$^{-1}$ was 12~mJy~beam$^{-1}$, after the primary beam correction.

\subsection{Moment maps: CO emission search with SoFiA}

We searched for CO emission using the Source Finding Application \citep[SoFiA; ][]{Serra:2015aa}.
We set up SoFiA to:
\begin{itemize}
\item normalize the cube by the local noise level in the mosaic; the noise varies by a factor of $\sim 5$ as a function of position on the sky (but only by a factor $1.4$ in the region where we detect CO emission) and by a factor $1.6$ along the frequency axis;
\item convolve the cube with a set of smoothing kernels, and build a detection mask which includes voxels outside the 4~$\sigma$ range in at least one of the convolved cubes; the smoothing kernels are circular Gaussians on the sky and box functions in velocity, and we use all possible combinations of Gaussian FWHM of 0, 3 pixels and box width of 0, 3, 7 channels;
\item construct individual objects by merging detected voxels with a friends-of-friends algorithm using a merging radius of 2 pixels along RA and Dec axes and 3 channels along the frequency axis;
\item remove from the detection mask all objects smaller than 3 pixels along RA or Dec axes or 3 channels along the frequency axis;
\item remove from the detection mask all objects with an integrated S/N below 3;
\item dilate the detection mask of the remaining objects by at most 2 pixels and 1 channel to include faint emission at the
edge of the detected objects.
\end{itemize}

The moment maps shown in Figure~\ref{fig:moment_maps} are obtained considering only voxels included in the resulting SoFiA detection mask.
In moment-1 and -2 images, we further blank pixels with a CO surface brightness below $0.9$~Jy~beam$^{-1}$~km~s$^{-1}$, which corresponds to 3~$\sigma$ when averaging 6 channels together ($3.0\times12\times10^{-3}$~Jy~beam$^{-1}\times9.9$~km~s$^{-1}\times\sqrt(6) \sim 0.9$~Jy~beam$^{-1}$~km~s$^{-1}$).
The continuum map in Figure~\ref{fig:moment_maps} was generated with the whole SPWs except for the $^{12}$CO($J$=1-0) and CN($N$=1-0) line-detected channels.
Note that the rest frequencies of nine hyperfine components of $N$=1$\rightarrow$0 of CN molecule are covered in the 113~GHz SPW and a channel range for a tentative detection at 112.85~GHz which probably corresponds to CN($N$=1-0, $J$=3/2-1/2, $F$=5/2-3/2) whose rest frequency of $113.499639$~GHz was excluded when generating continuum image\footnote{$<1\sigma$ for 9.9~km~s$^{-1}$ resolution data without smoothing and $\sim2\sigma$ for 9.9~km~s$^{-1}$ resolution data with a 7-ch boxcar smoothing.}.

\subsection{Keck/LRIS observations and data reduction}

We obtained optical spectra with LRIS \citep{Oke:1995aa,Rockosi:2010aa} installed at the 10-m Keck-I telescope on September~16, 2018 (UT, PI: S.~Perlmutter).
We used the 560 dichroic mirror.
The 600/4000 grism and the 400/8500 grating with a grating angle of 22.9~degree were used for the blue ($3100~\AA\lesssim \lambda \lesssim 5600~\AA$) and red channels ($5500~\AA\lesssim \lambda \lesssim 1~\mu$m), respectively.
With this set-up, multiple major atomic lines can be observed, such as \oii${\lambda 3727}$, H$\beta (\lambda 4861)$, \oiii${\lambda 5007}$, H$\alpha (\lambda 6563)$, and \nii$\lambda 6548,6583$.
The width of the slit is 1.0~arcsec.
These configurations provide spectral resolution of $R \equiv \lambda/\Delta \lambda \sim600-1,000$, corresponding to $\sim 300-500$~km~s$^{-1}$.
The pixel samplings are 2.18~\AA~pixel$^{-1}$ and 2.32~\AA~pixel$^{-1}$ in the blue and red channels, respectively.
The slit is set to cross the galaxy center with P.A. of 158.56~deg.
The exposure time is 480~sec.
We adopt 14 arcsec for the gap width between the two CCDs for both the blue and red channels\footnote{\url{https://www2.keck.hawaii.edu/realpublic/inst/lris/longslit\_geometry.html}}.
This is a rough value but does not change our conclusion.

The LRIS data are reduced as follows.
We first subtract bias images from the raw data and flat-field bias-subtracted data with domeflat spectra.
Wavelength is calibrated with sky emission lines for the red channel data.
In the blue channel, since the number of sky lines used for wavelength calibration is not large enough, lamp data is used.
Note that the telescope altitude when the lamp data is obtained is different from the one during the observation.
Recession velocity is calculated in the same definition as that in ALMA data.
Then, we add velocity offsets ($\sim-180$~km~s$^{-1}$) to the blue channel data so that \nii~(in the red channel) and \oii~(in the blue channel) emission line structures are consistent with each other.
Uncertainty in absolute wavelength calibration of the data is $\lesssim2$~\AA, roughly corresponding to an absolute value of velocities of $\sim100$~km~s$^{-1}$.
Therefore, we discuss only relative velocity differences between the different lines.

\section{Total mass, spatial distribution and velocity field of molecular gas}
\label{sec:results}

\begin{figure*}[t]
\begin{center}
\includegraphics[width=150mm, bb=0 0 360 297]{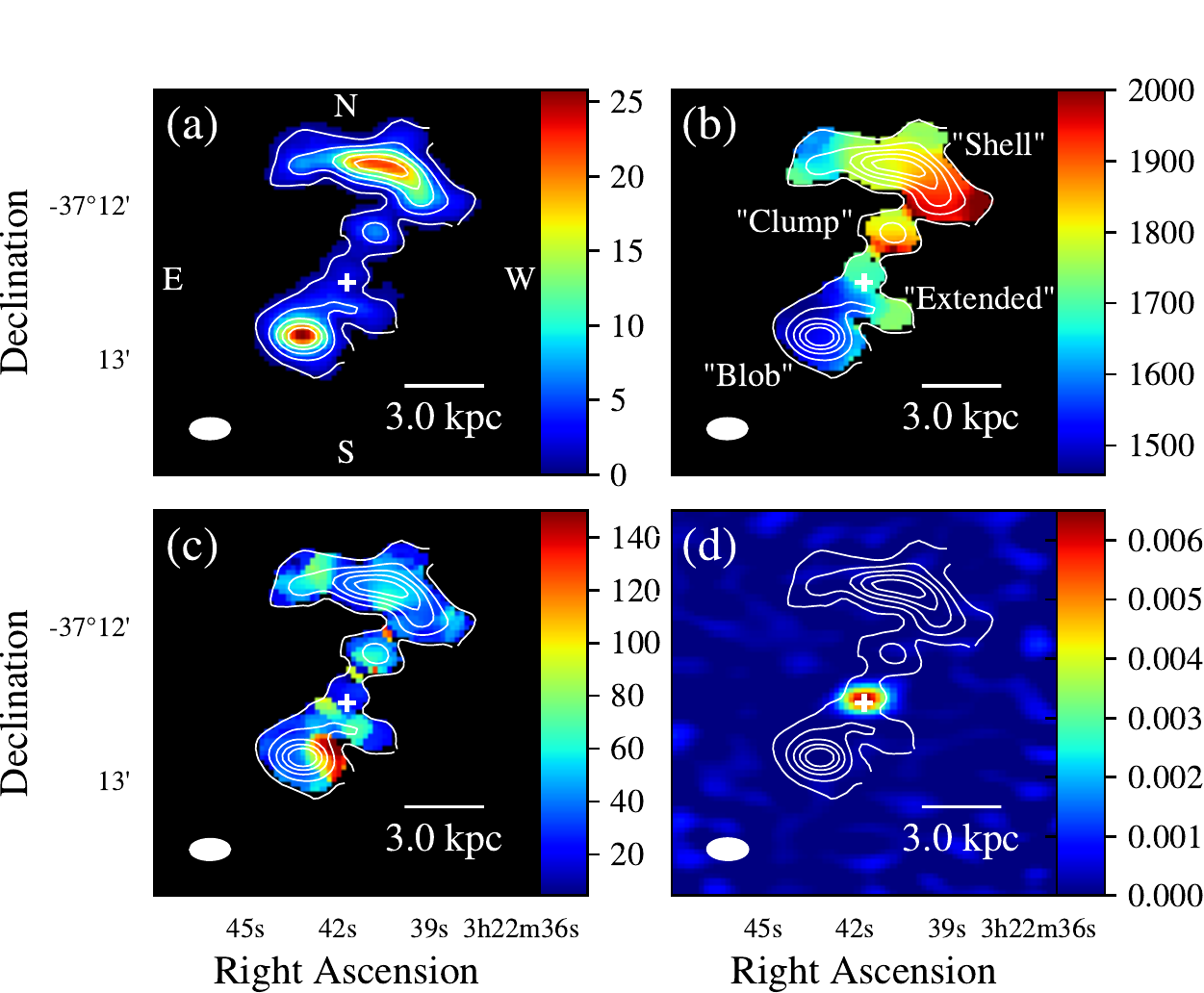}
\end{center}
\caption{Moment 0-2 and continuum maps of $^{12}$CO($J$=1-0) extracted from the data cube of NGC~1316 in the ICRS frame: (a) moment 0 (integrated intensity, Jy~beam$^{-1}$~km~s$^{-1}$), (b) moment 1 (velocity field, km~s$^{-1}$), (c) moment 2 (velocity dispersion, km~s$^{-1}$), and (d) continuum maps (mJy~beam$^{-1}$).
White contour shows moment 0 with contour levels of $[1.0, 5.0, 10, 20]\times 3\sigma_{\rm mom0,6ch}$, where $\sigma_{\rm mom0,6ch}$ is an rms noise of moment 0 map when averaging 6 channels together ($0.3$~Jy~beam$^{-1}$~km~s$^{-1}$).
Galaxy center is indicated with a white cross.
The resultant ALMA beam is shown as a filled white ellipse in the bottom left corner in each panel.
}
\label{fig:moment_maps}
\end{figure*}

\subsection{Total molecular gas mass}

The total molecular gas mass is estimated to be $(5.62 \pm 0.53)\times 10^8$~M$_\odot$ with the standard CO-to-H$_2$ conversion factor of the Milky-Way of $X_{\rm CO}=2.0\times 10^{20}$~cm$^{-2}$~(K~km~s$^{-1}$)$^{-1}$ \citep{Bolatto:2013rr}.
The conversion from Jy~beam$^{-1}$ to Kelvin is done based on the Rayleigh-Jeans approximation with an equation of $T=1.222\times10^6\frac{I}{\nu^2 \theta_{maj}\theta_{min}}$, where $T$ is the brightness temperature in Kelvin, $\nu$ is the observing frequency in GHz, $\theta_{maj}$ and $\theta_{min}$ are half-power beam widths along the major and minor axes in arcsec, respectively and $I$ is the brightness in Jy~beam$^{-1}$.

The molecular gas mass is calculated by summing the fluxes of the original moment-0 map (without any masks) for a frequency range of 114.497-114.73~GHz (or a channel range of 191-252 ch with velocity resolution of 9.9~km~s$^{-1}$ or a velocity range of 1406.8-2013.55~km~s$^{-1}$ with a rest frame frequency of $115.271202$~GHz) enclosed with a box centered on (R.A., Dec.)$_{\rm ICRS}=(03^{h}22^{m}41^{s}.448, -37^{\circ}12'16''.2)$ with a width of 120~arcsec and a height of 160~arcsec to cover the whole CO line-detected region.
The error for the molecular mass is calculated with a noise map as $\sigma_{rms} \sqrt{npts} ({\rm pixel\ scale})^2$, where $\sigma_{rms}$ and $npts$ are the rms and the number of pixels of the box used to calculate total flux, respectively.
The noise map is generated with CO line-free channels at the lower and higher frequency-sides of the CO channels.
Note that the number of channels used to generate the noise map is the same as the number of channels to calculate the moment~0.
The helium contribution to mass is accounted for by multiplying with a factor of $1.36$.
The obtained molecular gas mass is comparable to the previous single-dish observations \citep[$(4.0-5.4) \times 10^8$~M$_\odot$,][]{Wiklind:1989aa,Sage:1993aa,Horellou:2001aa}.

\begin{figure*}[t]
\begin{center}
\includegraphics[width=150mm, bb=0 0 489 357]{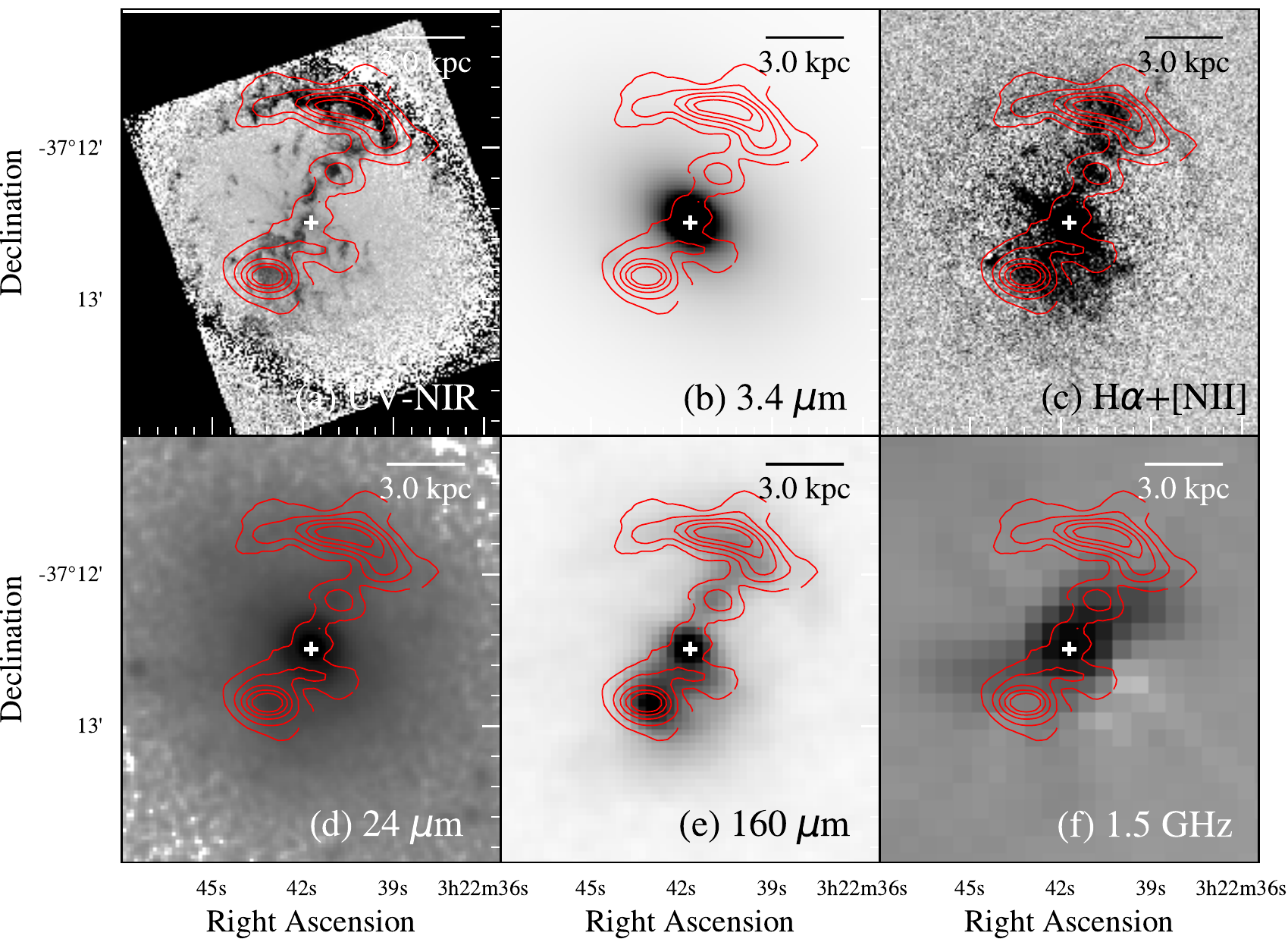}
\end{center}
 \caption{Comparison with other wavelengths data: (a) HST F336W-F160W, (b) WISE 3.4~$\mu$m, (c) CTIO H$\alpha$+\nii~\citep{Kennicutt:2003eu} (d) Spitzer/MIPS 24~$\mu$m \citep{Kennicutt:2003eu}, (e) Herschel/PACS 160~$\mu$m, and (f) VLA 1.5~GHz \citep{Fomalont:1989aa}.
 CO integrated intensity map is shown as a red contour on each map (same as Figure~\ref{fig:moment_maps}).
 CO distribution beautifully matches cold dust (a and e) and ionized gas (c) distributions.
 Note that the coordinates of the other wavelengths images are in the FK5 frame (J2000 equinox) whereas those of ALMA image is in the ICRS frame.}
\label{fig:comparison}
\end{figure*}

\subsection{Spatial distribution}
\label{sec:distribution}

\begin{table*}
  \tbl{Flux density for each component}{%
  \begin{tabular}{lccccc}
      \hline
      Name & Flux density & Molecular mass & Box center$^*$ & Width$^{**}$ & Height$^{**}$\\ 
       & [Jy] & [$10^7$~M$_\odot$] &  & [arcsec] & [arcsec]\\ 
      \hline
      Shell & $72.4 \pm 1.2$ & $33.5 \pm 0.6$ & $(3^{h}22^{m}41^{s}.002, -37^{\circ}11'46''.359)$ & $81.3445$ & $35.9328$\\ 
      Clump & $6.37 \pm 0.48$ & $2.95 \pm 0.22$ & $(3^{h}22^{m}40^{s}.805, -37^{\circ}12'11''.545)$ & $19.4958$ & $13.7687$\\ 
      Center & $1.92 \pm 0.35$ & $0.89 \pm 0.16$ & $(3^{h}22^{m}41^{s}.718, -37^{\circ}12'29''.62)$ & $14.7568$ & $9.8306$\\
      Extended & $6.55 \pm 0.47$ & $3.04 \pm 0.22$ & $(3^{h}22^{m}42^{s}.295, -37^{\circ}12'38''.563)$ & $41.3445$ & $7.38806$\\
      Blob & $34.3 \pm 0.5$ & $15.9 \pm 0.2$ & $(3^{h}22^{m}43^{s}.239, -37^{\circ}12'51''.844)$ & $31.2605$ & $19.1418$\\
      \hline
    \end{tabular}}\label{tab:flux}
\begin{tabnote}
\footnotemark[$*$] Center coordinate of the box to calculate flux density.\\
\footnotemark[$**$] A width and a height of the box to calculate flux density.
\end{tabnote}
\end{table*}

For the first time, our ALMA data clearly show the shell structure (hereafter ``Shell'', see Figure~\ref{fig:moment_maps}) in the northwest (NW) side and a large concentration (hereafter ``Blob'' in Figure~\ref{fig:moment_maps}) in the southeast (SE) side.
A small clump was detected along the line connecting the NW Shell and the SE Blob (``Clump'' in Figure~\ref{fig:moment_maps}).
There are also a weak emission at the center and an extended structure in the east-west direction just above the SE Blob (``Extended'' in Figure~\ref{fig:moment_maps}).
The fluxes and molecular gas mass for each component are summarized in Table~\ref{tab:flux}.
In Figure~\ref{fig:comparison}, we can see that the spatial distribution of molecular gas traced with $^{12}$CO($J$=1-0) excellently matches the dust patches visible in the optical \citep{Grillmair:1999aa} and FIR images \citep{Xilouris:2004aa,Temi:2005aa,Lanz:2010aa,Duah-Asabere:2016aa}, as suggested in the previous studies using lower spatial-resolution CO data \citep{Horellou:2001aa}.

\begin{figure}[t]
\begin{center}
\includegraphics[width=80mm, bb=0 0 405 305]{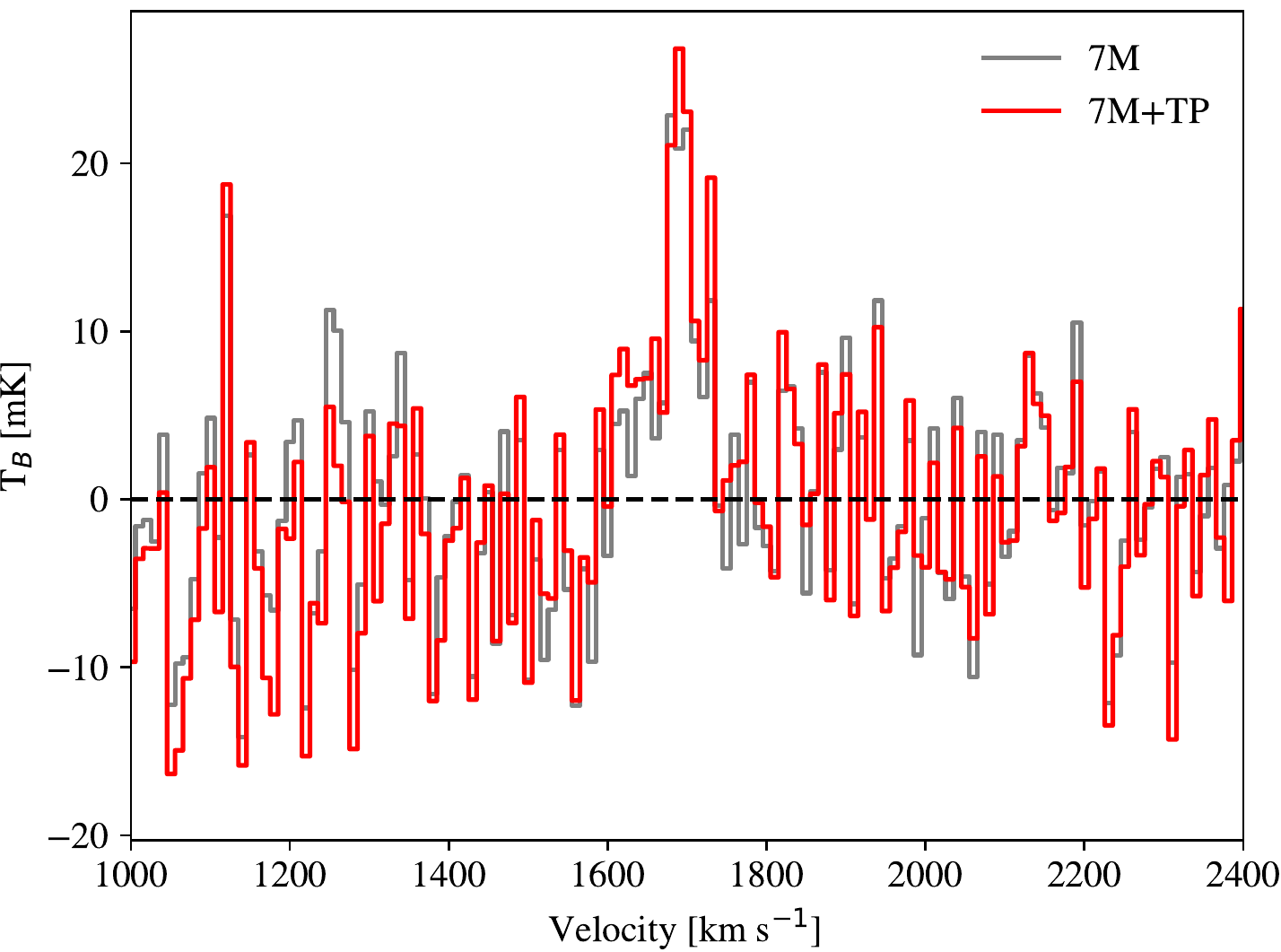}
\end{center}
\caption{CO spectra at the central position of 7M (grey) and 7M+TP data (red).
CO spectra are generated in the box area centered on the coordinate of NGC~1316 with a width of 15.18~arcsec (major axis of the beam) and a height of 7.73~arcsec (minor axis of the beam).
CO emission is detected at a velocity of $\sim1720$~km~s$^{-1}$ with a confidence level of $4.2\sigma$.
}
\label{fig:center_spectra}
\end{figure}

At the central position, a weak and narrow CO emission is detected with a peak intensity of $26.8$~mK at $\sim 1720$~km~s$^{-1}$ in our $9.9$~km~s$^{-1}$ velocity-resolution data ($4.2 \sigma$, Figure~\ref{fig:center_spectra}), which is consistent within the error with the systemic kinematic-LSR velocity in {\it radio} definition of $1732\pm10$~km~s$^{-1}$, which is measured with optical spectroscopy \citep{Longhetti:1998aa}.
\cite{Horellou:2001aa} reported a very broad emission ($\sim 500$~km~s$^{-1}$ at the base) at the central position with a $43''$ beam.
As they stated, the CO line width is broad because their $43''$ beam encloses parts of the NW clump and the SE extended structure.

\subsection{Velocity field: not only a simple rotating disk?}
\label{sec:velocityfield}

In this section, we introduce the velocity field of molecular gas in section~\ref{sec:co_vel} and compare with the velocity field of ionized gas obtained with our Keck/LRIS observations in section~\ref{sec:compare_vel}.

\subsubsection{CO velocity field}
\label{sec:co_vel}

\begin{figure*}[h]
\begin{center}
\includegraphics[width=170mm, bb=0 0 701 253]{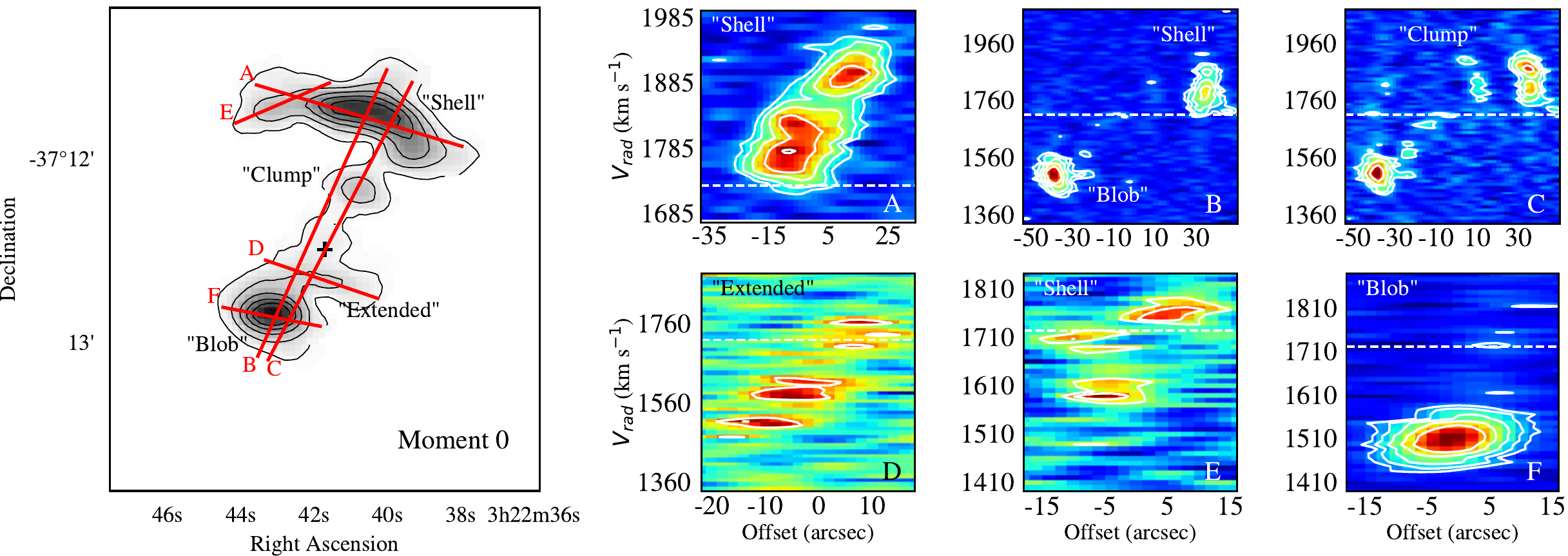}
\end{center}
 \caption{Position-Velocity (PV) maps sliced with red lines shown on the CO integrated intensity map on the left-hand side.
 The starting point of each slice (the smallest offset value) is indicated with an alphabet of each PV map.
 Black cross indicates the central position of NGC~1316 in the moment-0 map.
 Dashed lines on PV maps indicate the systemic velocity of NGC~1316 of 1732~km~s$^{-1}$ \citep{Longhetti:1998aa}.
 Contour levels are $[3, 5, 8, 12, 16]\times \sigma_{rms}$, where $\sigma_{rms}$ is an rms noise of $12$~mJy/beam.
 ALMA observations revealed very complex velocity field of NGC~1316, for example, two velocity components at offset of 10~arcsec in slice A (north-east side), and not a simple interpolation between the NW Shell and SE Blob (C), and two velocity components in slice E.
 }
\label{fig:pv_maps}
\end{figure*}

The high spatial resolution and sensitivity of the ALMA data reveals a very complex velocity structure which cannot be explained {\it only} by a ``polar'' rotating disk with kinematical major axis along P.A.~$\sim -20$~deg (the axis along which the CO velocity range is the largest).
Figure~\ref{fig:moment_maps}~(b) and (c) show the moment-1 and moment-2 maps, respectively.
Our ALMA data clearly show a monotonic trend in velocity along the NW Shell, $\sim1600$~km~s$^{-1}$ at the east edge and $\sim2000$~km~s$^{-1}$ at the west edge.
This is not what we expect for a rotating disk, i.e, spider diagram, where the velocity increases (decreases) when moving towards the kinematical major axis and decreases (increases) when moving away from it for the receding (approaching) side.

The velocity dispersion of the molecular gas is roughly in the range of $20-50$~km~s$^{-1}$ at kpc-resolution.
There are some areas with as high velocity dispersion as $>80$~km~s$^{-1}$ and a very high value of $\sim150$~km~s$^{-1}$ is found at the west edge of the SE Blob.
The former area is an overlapping region of the Shell ($\sim2000$~km~s$^{-1}$) and Clump ($\sim1740$~km~s$^{-1}$) components.
The latter is due to multiple velocity components around $\sim1530$~km~s$^{-1}$, $\sim1730$~km~s$^{-1}$ and $\sim1814$~km~s$^{-1}$ (Figure~\ref{fig:pv_maps}, ``Offset'' of 5 arcsec in the PV cut ``F'').
Again, this trend in the moment-2 map is not what we expect for a rotating disk.
We often see a high velocity dispersion along the kinematical major axis, since there is an abrupt change in the direction of the velocity vector of gas within the beam along the axis.

In Figure~\ref{fig:pv_maps}, the position-velocity (PV) maps are presented.
The velocity difference between the NW Shell and the SE Blob ($\sim 400$~km~s$^{-1}$, see PV cut ``B'', a line connecting the intensity peaks of NW Shell and SE Blob) is consistent with the value reported in previous CO studies \citep{Horellou:2001aa,Lanz:2010aa}.
The followings are new findings with our ALMA data.
The PV cut ``C'' (a line passing through the Shell, Clump, the galaxy center, Extended, and Blob) suggests that some components seem to follow regular rotation but others do not (e.g., ``Clump'').
This system may be a combination of a (nearly edge-on) rotating disk/ring and components with disturbed kinematics.
The relative amount of the disturbed gas to the rotating gas is large, which is contrary to the other radio galaxies with dust or molecular gas \citep{Kotanyi:1979aa,de-Koff:2000aa,de-Ruiter:2002aa,Ruffa:2019aa}.

In the NW Shell, there are two velocity components ($\sim1800$~km~s$^{-1}$ and $\sim 1875$~km~s$^{-1}$) near the brightest point (PV cuts along ``B'' and ``C'').
In addition, there are at least two velocity components at the east edge of the NW Shell.
We can see along the PV cut ``A'' and ``E'' that the NW Shell spans both velocities below and above systemic velocity, which is also one of the features unlikely for a rotating disk.

The SE Blob was barely resolved with our beam and likely to have velocity gradient of $\sim 125$~km~s$^{-1}$ along northeast to southwest direction (PV cut ``F'').
In the SE Blob, there are multiple velocity components, the main body around $1500$~km~s$^{-1}$, and very weak components around $1650$~km~s$^{-1}$, $1725$~km~s$^{-1}$, and $1825$~km~s$^{-1}$ (``Offset'' of around $-35$ arcsec in the PV cut plot along ``C'' in Figure~\ref{fig:pv_maps}).
With the sensitivity of our data, it seems that the $1725$~km~s$^{-1}$ and $1825$~km~s$^{-1}$ components are discrete structures but the $1650$~km~s$^{-1}$ component is physically connected with the main body of the SE Blob.

The ``Extended'' component above the SE Blob has a continuous velocity gradient, $\sim 1475$~km~s$^{-1}$ at the east-side edge and $\sim 1775$~km~s$^{-1}$ at the west-side edge (PV cut ``D'').
The ``Clump'' seems to have a velocity gradient in the north-south direction (PV cut ``C'' and Figure~\ref{fig:moment_maps}b).
These CO components revealed with ALMA likely correspond to the dust patches recognized in ultraviolet, optical \citep[Figure~\ref{fig:comparison}a or Figure~5 of][]{Iodice:2017aa} and FIR data (Figure~\ref{fig:comparison}e or Figures~2, 5 of \citealt{Lanz:2010aa}; Figure~5 of \citealt{Duah-Asabere:2016aa}).

\subsubsection{Comparison between CO and ionized gas}
\label{sec:compare_vel}

\begin{figure}[h]
\begin{center}
\includegraphics[width=80mm, bb=0 0 391 306]{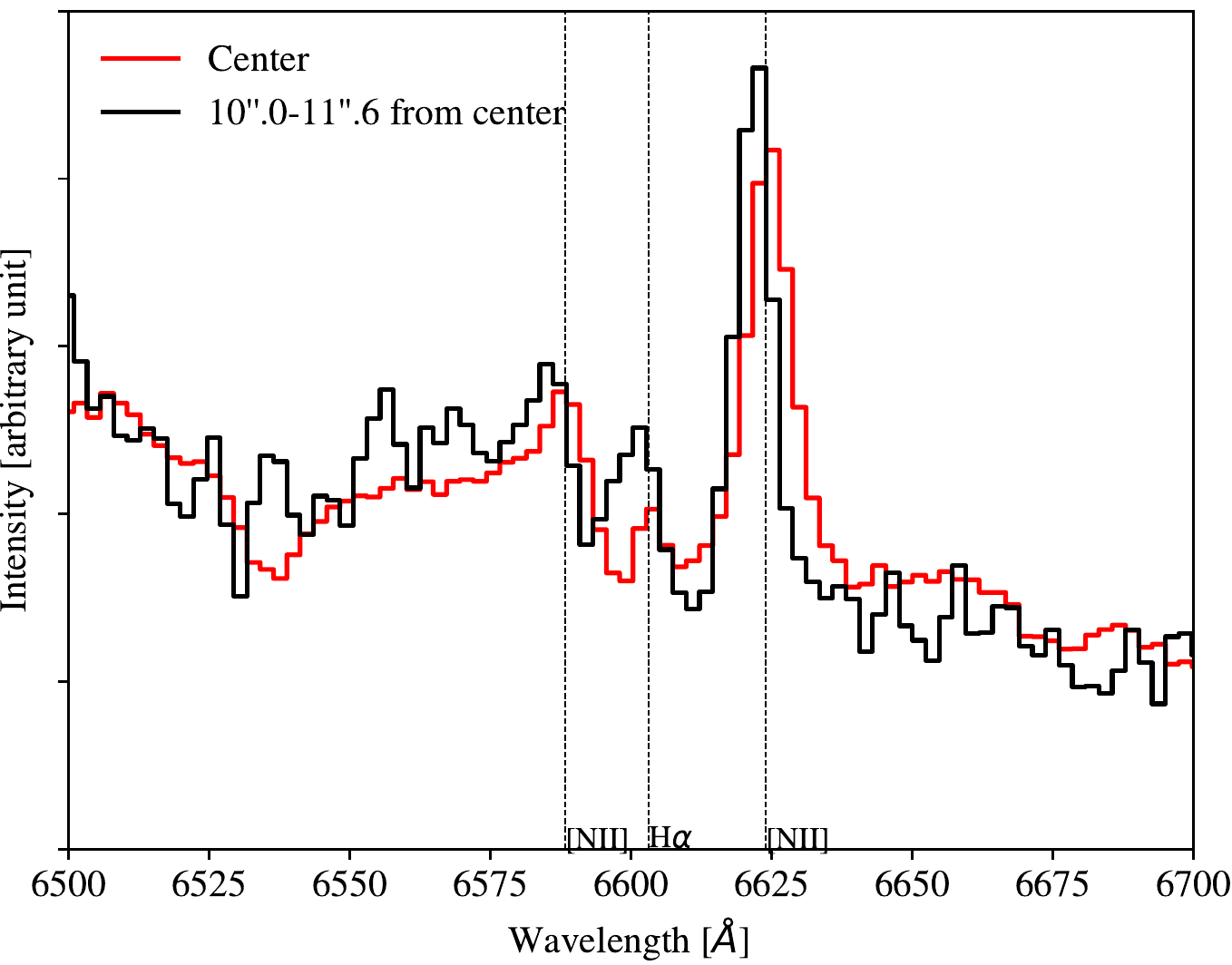}
\end{center}
 \caption{Keck/LRIS spectra around the wavelength of H$\alpha$ and \nii~at the galaxy center (red) and at $10''.0-11''.6$ from the center (black) where the strongest H$\alpha$ emission is found.
 Vertical lines indicate the redshifted H$\alpha$ and \nii~with the galaxy redshift ($z=0.00587$).
 H$\alpha$ is detected but very weak.
 The emission lines at $10''.0-11''.6$ from the center are slightly shifted to shorter wavelengths due to the gas dynamics.
 }
\label{fig:ha}
\end{figure}

\begin{figure}[h]
\begin{center}
\includegraphics[width=80mm, bb=0 0 225 291]{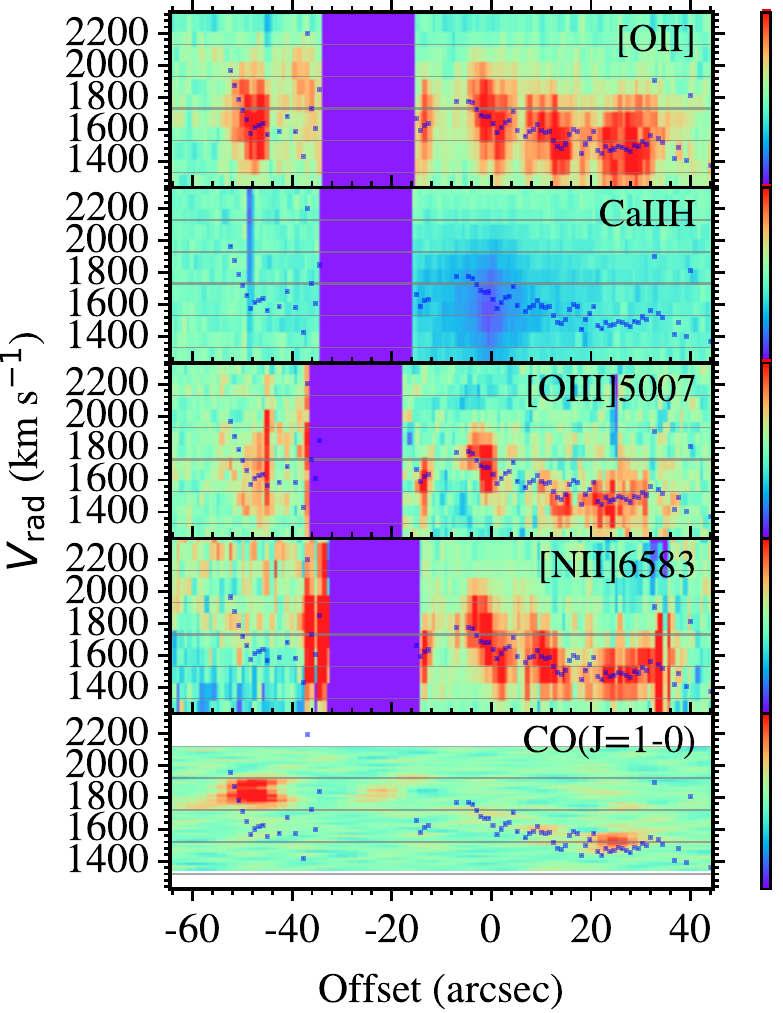}
\end{center}
 \caption{PV maps of CO, ionized gas (\oii,  \oiii, and \nii), and stellar absorption (\caii~H) obtained with Keck/LRIS observations (P.A. of the LRIS slit is $158.56^\circ$, or the PV cut C with a longer slice length in Figure~\ref{fig:pv_maps}).
 Intensities are arbitrally scaled so as to be easily seen.
 The black dots indicate the intensity peak of the strongest line, \oii.
 The horizontal lines indicates the relative velocities with respect to the systemic velocity, $1732\pm n \times 200$~km~s$^{-1}$ where $n=0,1,2$.
 Purple-shaded areas indicate the gap between two CCDs.
 The velocity of the ionized gas is roughly consistent with that of molecular gas, despite some disagreements.
 }
\label{fig:pv_maps_keck}
\end{figure}

We found that \oii, \oiii~and \nii~are detected in emission and \caii~HK are detected in absorption.
Figure~\ref{fig:ha} shows H$\alpha$ and \nii~spectra where H$\alpha$ emission seems to be detected.
H$\alpha$ is detected but very weak maybe due to it is a combination of the emission of ionized gas and stellar absorption.
Therefore, Figure~\ref{fig:comparison}c has been supposed to be an H$\alpha$+\nii~image, but the contribution from \nii~is likely to be more dominant than H$\alpha$.

The PV maps of \oii, \caii~H, \oiii, and \nii~are presented in Figure~\ref{fig:pv_maps_keck}.
Note that the intensity of each line in this figure has not been calibrated. 
The \caii~H absorption line is not resolved in our data with velocity resolution of $300-500$~km~s$^{-1}$ and its velocity is almost constant around the galaxy center.
On the other hand, the other lines show a velocity gradient, $\sim1500$~km~s$^{-1}$ at the offset of 25~arcsec from the galaxy center (SE) and $\sim1900$~km~s$^{-1}$ at $-50$~arcsec offset (NW).
Although the velocity resolution is as coarse as $300-500$~km~s$^{-1}$, the accuracy of determining the central velocity at each pixel is estimated to be $\lesssim100$~km~s$^{-1}$.
Note that \cite{Schweizer:1980aa} reported a larger velocity gradient of $\sim 700$~km~s$^{-1}$ but with a different slit angle (P.A. of 142$^\circ$).
There also exist some irregularities which cannot be explained only by rotation, such as a velocity depression at $-15$~arcsec.

Our optical data showed that the kinematics of the ionized gas cannot be explained only by rotation, as we found in the molecular gas data, or the kinematic major axis is largely different from the slit angle of $158.56^\circ$.
Though the velocity resolution of the ALMA and Keck data are very different ($\sim10$~km~s$^{-1}$ vs $\sim100$~km~s$^{-1}$), the velocity gradient of the molecular gas at the same P.A. is roughly consistent with the one seen in ionized gas (Figure~\ref{fig:pv_maps_keck}).
The systemic velocity of $1732\pm10$~km~s$^{-1}$ recorded in the NED (Table~\ref{tab:first}) is derived from the central \oii~velocity \citep{Longhetti:1998aa} and consistent with our measurement.
However, there is a significant offset of a few tens~km~s$^{-1}$ in the velocity of the stellar absorption \caii~and this is considered to be a real systemic velocity of NGC~1316.
It is important to obtain integral field unit (IFU) data, compare the velocity fields of ionized gas and molecular gas in more detail, and also investigate the excitation mechanism of the ionized gas.


\section{Implications for the kinematic nature of the molecular gas}
\label{sec:discussions}

In this section, we discuss the kinematic nature of the molecular gas observed in the central region of NGC~1316: inflow (section~\ref{sec:external}) or outflow (section~\ref{sec:internal}).

\subsection{Relative line-of-sight location of each component}

\begin{figure*}[h]
\begin{center}
\includegraphics[width=150mm, bb=0 0 900 264]{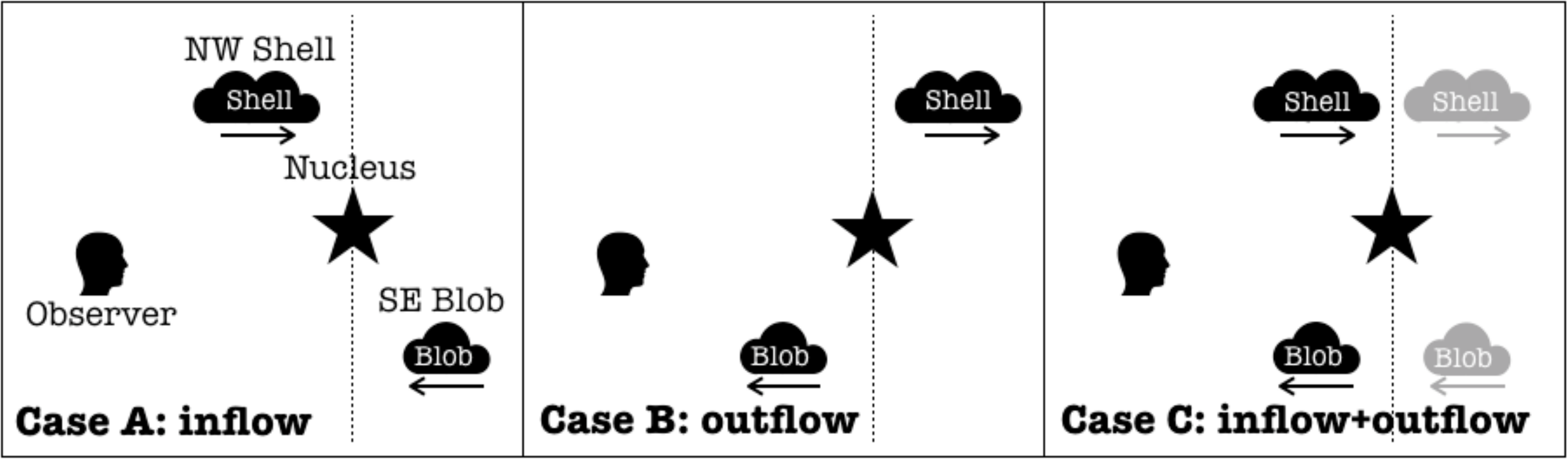}
\end{center}
 \caption{Schematic view of three cases of the line-of-sight location of the NW Shell, SE Blob and Nucleus of NGC~1316.
 The arrows indicate the motion direction of the molecular gas which is not following regular rotation with respect to the nucleus of NGC~1316.
 For the case A, most of the gas in the NW is between us and the galaxy center (or the radio AGN), and the gas deviating from rotation is likely inflowing.
 For the case B, most of the gas in the SE is in between us and the galaxy center (or the radio AGN), and the gas deviating from rotation is likely outflowing.
 }
\label{fig:schematic}
\end{figure*}

Once the relative line-of-sight location of the NW Shell, the SE Blob, and the nucleus (``Nucleus'', marked with a cross in Figure~\ref{fig:moment_maps}) is determined (Figure~\ref{fig:schematic}), more detailed kinematic information of the molecular gas can be derived.
Our ALMA observations revealed that larger recession velocity of the NW Shell ($\sim 1850$~km~s$^{-1}$) and smaller recession velocity of the SE Blob ($\sim 1500$~km~s$^{-1}$) than the systemic velocity of NGC~1316 \cite[1732~km~s$^{-1}$,][]{Longhetti:1998aa}.
Assuming gas in this region is not rotating, the case of Shell-Nucleus-Blob order from the observer suggests inflow motion (case A)
the Blob-Nucleus-Shell case suggests outflow motion (case B)
and the cases where both the NW Shell and SE Blob are located on the same side with respect to Nucleus suggest a coexistence of inflow and outflow motions (case C).

Our ALMA data and the ancillary data seems to favor the case A.
In Figure~\ref{fig:comparison}, both the CO and 160~$\mu$m dust emission are stronger at the peak of SE Blob than that of NW Shell (panel (e)), whereas the dust extinction is stronger at the NW shell than the SE Blob (panel (a)) and the distribution of stellar component is smooth (panel (b)).
This suggests that there is a larger amount of ISM at the SE Blob than at the NW Shell whereas the extinction by the SE Blob is weaker than the NW Shell.
However, $160\mu$m brightness does not only reflect the amount of dust but also the dust temperature and the strength of the back ground far-ultraviolet radiation field, i.e., heating source of the dust.
Therefore we consider the case A most plausible, while we leave the possibility of the other cases in the following discussions.
The inclination of the nuclear jet would provide us further hints on the relative line-of-sight location of the Shell, the Blob, and the Nucleus but it is difficult to estimate it with our current data set.

\subsection{Inflow?}
\label{sec:external}

The previous studies on NGC~1316 consider that the dust and molecular gas of the galaxy is injected via wet minor mergers \citep[e.g.,][]{Sage:1993aa,Horellou:2001aa,Lanz:2010aa}, since ETGs are generally gas-poor.
Generally, the gas-rich ETGs tend to show disturbed morphology in deep optical images \citep{van-Dokkum:2005aa,Duc:2015aa}, suggesting recent merging events as observed in NGC~1316.
\cite{Lanz:2010aa} estimated the dust mass of NGC~1316 to be $2.0 \times 10^{7}$~M$_\odot$ (corrected for the different choice of the distance) from $24$~$\mu$m, $70$~$\mu$m, and $160$~$\mu$m data taken with the Multiband Imaging Photometer (MIPS) on the Spitzer space telescope.
They showed that the dust-to-stellar mass ratio of NGC~1316 is $>100$ times larger than the value expected from the empirical relation between the dust-to-stellar mass ratio and stellar mass of elliptical galaxies and concluded that the dust currently present in NGC~1316 was injected by a merger galaxy.
They also claimed that the spatial coincidence between dust and CO suggests a common origin for them.

In the case of A or C as described above (Figure~\ref{fig:schematic}), the molecular gas is likely (at least partially) inflowing.
Especially, the multi-wavelength data of NGC~1316 seem to favor the case A.
In this section, we discuss the infalling galaxy (section~\ref{sec:infallinggalaxy}) and its merging timescale (section~\ref{sec:timescales}).

\subsubsection{Infalling galaxy (galaxies)}
\label{sec:infallinggalaxy}

For NGC~1316, \cite{Lanz:2010aa} estimated the stellar mass of the merged galaxy to be $(1-6)\times 10^{10}$~M$_\odot$ which injected the dust and gas to the central region based on the dust mass and the dust-to-stellar mass ratios of Sa-Sm galaxies (assuming that the merged galaxy is an Sa-Sm galaxy).

Stellar masses of infalling galaxies can be also estimated with a gas-to-dust ratio (GDR) and an H$_2$/\HI~ratio combined with an empirical relations between stellar mass and GDR and H$_2$/\HI~ratio.
\cite{Davis:2015aa} found a high GDR of $2.0-730$ (the median of 315) and a low H$_2$/\HI~ratio of $0.1-2.0$ (the median of 0.4) in ETGs with prominent dust lanes (corrected for the use of different $\alpha_{\rm CO}$), suggesting a recent merger with a lower mass companion (predicted stellar mass ratio of $40:1$).
Note that the stellar masses of \cite{Davis:2015aa} sample are $10^{10.5-11.2}$~M$_\odot$ (the median of $10^{10.8}$~M$_\odot$), which is smaller than that of NGC~1316 of $10^{11.8}$~M$_\odot$.
One of the nearby ETGs with a prominent dust lane, NGC~5128 (Centaurus~A) whose stellar mass of $1.1\times10^{11}$, has a GDR of $103$ \citep{Parkin:2012aa} and an H$_2$/\HI~ratio of $1.06$ \citep{Charmandaris:2000aa}.

For NGC~1316, given the literature data of dust mass \citep[$2.0\times10^{7}$~M$_\odot$,][]{Draine:2007aa} and the upper limit of \HI~mass \citep[$<10^8$~M$_\odot$,][]{Horellou:2001aa}, the obtained GDR and H$_2$/\HI~ratio of the dust patches are respectively $\sim 28$ and $\gtrsim 5.6$, which are roughly $\sim10$ times lower and higher than the median values found in the ETGs with dust lane.
The inferred stellar mass of accreted galaxy to NGC~1316 using the empirical relation of these ratios and stellar mass is $>10^{11.5}$~M$_\odot$ \citep{Remy-Ruyer:2014aa,Bothwell:2014lr,Davis:2015aa}.
Considering an ordered stellar distribution of the main body of NGC~1316 \citep{Schweizer:1980aa,Iodice:2017aa}, it is unlikely that NGC~1316 is {\it now} experiencing a major merger.
In the following paragraphs, we discuss three possible explanations on the observed low GDR and high H$_2$/\HI~ratio of NGC~1316.

First, it is possible that the atomic gas of the infalling galaxy had been selectively stripped by ram pressure from hot halo of NGC~1316.
\cite{Cortese:2016aa} found that \HI-deficient cluster galaxies are poorer in atomic but richer in molecular hydrogen if normalized to their dust content.
This trend can be considered as a consequence of the selective stripping of the components distributed at the outer radius of galaxies by ram pressure of the inter-galactic hot gas.
\cite{Cortese:2016aa} showed that the low GDR and high H$_2$/\HI~ratio are reproduced with a numerical simulations of ram-pressure gas stripping with models of \cite{Bekki:2013aa,Bekki:2014aa,Bekki:2014ab}.
If we assume that only atomic gas of the infalling galaxy was selectively and completely stripped by ram pressure from hot halo of NGC~1316 and H$_2$/\HI~ratio of the progenitor was 0.1 (0.01) as dwarf galaxies, the GDR and H$_2$/\HI~ratio of the infalling galaxy are used to be $\sim 280$ ($\sim 2800$) and $\sim 0.56$ ($\sim 0.056$), respectively.
These ratios correspond to galaxies with stellar masses of $\sim 10^{9}$~M$_\odot$ ($< 10^{7}$~M$_\odot$) and $\sim 10^{10.5}$~M$_\odot$ ($\sim 10^{9}$~M$_\odot$), respectively.
However, it should be noted that NGC~1316 is located at far from the center of the Fornax cluster and its own hot halo is not significantly prominent \citep{Feigelson:1995aa,Kaneda:1995aa,Tashiro:2001aa,Isobe:2006aa,Tashiro:2009aa,Seta:2013aa}.
In addition, NGC~1316 is considered to be now infalling onto the Fornax cluster for the first time \citep{Drinkwater:2001aa}.
Therefore, the ram pressure stripping alone may not be able to explain the observed low GDR and high H$_2$/\HI~ratio of NGC~1316.

Second, it is claimed that the dust formation and conversion from atomic gas to molecular gas are enhanced in galaxy interactions \citep{Young:1989aa,Sanders:1996aa,Nakanishi:2006zv,Kaneko:2017aa}.
\cite{Sage:1993aa} discussed that the high H$_2$/\HI~ratio of their sample ETGs including NGC~1316 may be due to interaction or infall events. 
Conversion from atomic to molecular gas occurs primarily on the surface of dust grains \citep{Hollenbach:1971aa,Cazaux:2004aa} down to metallicities as low as $\sim 10^{-5}$ of solar value \citep{Omukai:2010th}.
Once the starburst is triggered due to the galaxy interaction, the dust is formed and blasted into interstellar space via core-collapse supernovae within the lifetime of massive stars \citep[e.g.,][]{Cernuschi:1967aa,Hoyle:1970aa,Gall:2014aa}.
In this case, the timescale of the conversion, $t_{\rm chem}$ is calculated as $t_{\rm chem} = \frac{1}{n_{\rm H} R_{\rm H_2}}$,
where $n_{\rm H} ({\rm cm^{-3}})$ is the number density of H atoms, and $R_{\rm H_2}\ ({\rm cm^3\ s^{-1}})$ is the rate constant of H$_2$ formation.
$R_{\rm H_2}$ is reported to be roughly in the range of $10^{-17}-10^{-16}\ {\rm cm^3\ s^{-1}}$ \citep{Jura:1975aa,Gry:2002aa,Browning:2003aa,Habart:2003aa}.
If we adopt a typical volume density of cold neutral medium of $\sim 10~{\rm cm^{-3}}$ as $n_{\rm H}$, $t_{\rm chem}$ becomes $\sim 3 \times 10^{7}$~yr.
Considering that the density should increase during the galaxy merger, this is an upper limit.
This is smaller than the dynamical time of the system ($3\times10^8$~yr, see Section~\ref{sec:timescales}) so this process is likely to be occurred.

Third possibility is the ionization of atomic gas by AGN-related radiation \citep{Horellou:2001aa} or a shock produced by jet.
It is known that there is a currently low-activity AGN with low-ionization nuclear emission-line region (LINER) with a low-power nuclear radio jet in the central region of NGC~1316.
Based on the comparison between NGC~5128 and NGC~1316, \cite{Horellou:2001aa} discussed that the high H$_2$/\HI~ratio may be due to stronger effects from the nuclear activity in NGC~1316, e.g., brighter in X-ray for NGC~1316 than NGC~5128 and a perpendicular jet to the dust lane in NGC~5128 \citep{Espada:2009aa} but not in NGC~1316.
The ionization timescale is generally much shorter than the dynamical time of galaxies.
The fractional ionization is determined by ionization rate and recombination rate.
The recombination timescale ($t_{\rm r}$) is calculated as $t_{\rm r}=\frac{1}{n_{\rm e} \alpha_{\rm A}}$ \citep[where $n_{\rm e}$ is the volume electron density and $\alpha_{\rm A}$ is the total recombination coefficient,][]{Avrett:1988aa}.
A typical $t_{\rm r}$ is $\sim 10^5/n_{\rm e}$~yr.
Considering a typical $n_{\rm e}$ value of $10^8 {\rm cm}^{-3}$ in ``broad-line region (BLR)'', $10^3-10^6 {\rm cm}^{-3}$ in ``narrow-line region (NLR)'', $t_{\rm r}$ is much smaller than the galactic timescales ($3\times10^8$~yr, see Section~\ref{sec:timescales}).
In case of the cocoon in Cygnus~A, $n_{\rm e}\sim10^{-3}-10^{-1} {\rm cm}^{-3}$ \citep{Snios:2018aa} and $t_{\rm r}$ becomes $10^{6-8}$~yr.
However, it is unlikely that this is the only case, since it is unclear why the ionization of atomic gas is enhanced while the dissociation of molecular gas to atomic gas is not enhanced.
It may be possible that the dust shielding works effectively at the area where the molecular gas exist (large $A_V$).
In addition, if the most atomic gas is ionized, strong H$\alpha$ emission is expected to be observed, but no prominent H$\alpha$ emission was detected in our Keck/LRIS observations, as previous optical study suggested \citep{Schweizer:1980aa}.
It is necessary to observationally investigate the properties of the photo-dissociation region (PDR) of the central region of NGC~1316.

In summary, the second possibility (effective dust formation and conversion from atomic to molecular gas) seems to be a most plausible reason for the observed low GDR and high H$_2$/\HI~ratio of NGC~1316, and the other two mechanisms (ram-pressure stripping and ionization of atomic gas) may partially contribute to these unusual ratios.


\subsubsection{Timescales}
\label{sec:timescales}

\cite{Lauer:1995aa} considered NGC~1316 is in the earliest stage of the ``settling sequence'' of dust in the ETGs \citep{Tran:2001aa}.
\cite{Goudfrooij:2001aa,Goudfrooij:2001ab} estimated merger age to be $\approx 3$~Gyr by measuring the age distribution of globular clusters in NGC~1316 with Hubble Spacec Telescope (HST).
\cite{Sesto:2016aa,Sesto:2018aa} also measured the age of star clusters and found a younger cluster population ($\sim1-2$~Gyr) in addition to the intermediate-age clusters ($\sim5$~Gyr) which is the dominant population.
\cite{Lanz:2010aa} estimated a lower limit on the merger age to be 22~Myr by calculating the free-fall time of the central molecular gas to the galaxy center.

High-resolution ALMA data of NGC~1316 revealed a complex spatial distribution and velocity field of molecular gas, suggesting that the molecular gas has not been settled into a steady state yet, which has been already claimed in the previous dust extinction observations \citep{Lauer:1995aa}.
Theoretically predicted timescale for gas to settle into a disk, $t_{\rm relax}$ ranges from 10$^8$~yr to Hubble time depending on models, and a typical value is $10^9$~yr \citep[e.g.,][]{Gunn:1979aa,Tubbs:1980aa,Tohline:1982aa,Steiman-Cameron:1988aa,Habe:1985aa,Christodoulou:1992aa,Christodoulou:1993aa}. 
$t_{\rm relax}$ in theoretical models depends mainly on four factors \citep{Tohline:1982aa, Christodoulou:1992aa, West:1994aa}: (1) the degree of misalignment between the angular momentum vector of the gas disk and the symmetry axis of the galaxy; (2) the shape of the galactic potential; (3) the viscosity of the gas and the efficiency of the dissipative force, and (4) the distance from the galactic center.
$t_{\rm relax}$ is expected to be longer for larger misalignment for (1), more spherically symmetric potential for (2), smaller viscosity for (3), and larger distance for (4).

\cite{Lake:1983aa} investigated the star (collisionless component) and gas (dissipative component) orbits in triaxial potential and provided a formula to estimate $t_{\rm relax}$ with a dynamical time, $t_{\rm dyn}$ and an eccentricity of the potential, $\epsilon$ as $t_{\rm relax}=t_{\rm dyn}/\epsilon$.
They considered that the accreted and disrupted gas is organized into a sequence of tube orbits, and the tube orbits settled to the equatorial plane via a differential precession damped by a dissipative nature of gas.
This process is expected to occur on the timescale of the differential precession of the orbits.
For NGC~1316, $t_{\rm relax}$ is estimated to be $8\times10^8$~yr with $t_{\rm dyn}$ of $3\times10^8$~yr \citep[$\sim 100$~km~s$^{-1}$ at $5$~kpc,][]{Bosma:1985aa,DOnofrio:1995aa,Arnaboldi:1998aa,Bedregal:2006aa} and $\epsilon$ of 0.4 \citep[Fig.~7 of][]{Iodice:2017aa}.
Here, we assume that the eccentricity of the potential is the same as that of stellar component.
Note that the timescale is an upper limit since gas viscosity is not taken into account.
However, the effect of the viscosity on $t_{\rm relax}$ is claimed to be as small as $t_{\rm relax}\propto$~(viscosity coefficient)$^{-1/3}$ \citep{Steiman-Cameron:1988aa}.
\cite{Varnas:1990aa} showed that $t_{\rm relax}$ is changed by a factor of $\sim2$ when changing the viscosity parameter by a factor of 10.
Therefore, our data suggest that the central molecular gas is injected within $\sim 1$~Gyr.

The $t_{\rm relax}$ of $\sim 1$~Gyr is also supported by a gas depletion time by star formation, $t_{\rm dep}$ of NGC~1316.
$t_{\rm dep}$ is defined as a total cold gas mass divided by star formation rate (SFR).
If the $t_{\rm dep}$ is much shorter than the $t_{\rm relax}$, it is unlikely that the molecular gas survive for $t_{\rm relax}$.
SFR of NGC~1316 is estimated to be $0.30-0.77$~M$_\odot$ yr$^{-1}$ with total infrared luminosity that an AGN contribution is not considered \citep{Duah-Asabere:2016aa}.
Thus, $t_{\rm dep}$ of NGC~1316 is calculated to be $\sim 1$~Gyr or longer, which is comparable to local late-type galaxies \citep[e.g.,][]{Kennicutt:1998hb,Bigiel:2011uq,Kennicutt:2012yq}.
The star formation properties of NGC~1316 will be investigated in the forthcoming paper (Morokuma-Matsui et al.~in prep.).



However, it is possible that the $t_{\rm relax}$ for NGC~1316 is longer than $\sim1$~Gyr.
\cite{Davis:2016aa} claimed that a longer $t_{\rm relax} \sim 100~t_{\rm dyn} \approx 1-5$~Gyr is required to explain the observed histogram of the difference between the projected angular momenta of stellar and molecular gas of ETGs.
Numerical simulations showed that $t_{\rm relax}$ becomes as long as Hubble time under the spherically symmetric potential \citep{Christodoulou:1992aa}.
We assumed here that the eccentricity of potential is the same as that of stellar component but it is non-trivial that they are the same.
In addition, the true SFR (without AGN contribution) may be much smaller than the adopted value for NGC~1316.
In order to constrain the timescale of the merger event for NGC~1316, it is needed to conduct numerical simulations with dust and molecular gas formation as well as AGN feedbacks.



\subsection{Outflow?}
\label{sec:internal}

There is a number of galaxies with AGN (jet)-driven outflows of molecular gas \citep[e.g.,][]{Alatalo:2011aa,Combes:2013ab,Cicone:2014aa,Garcia-Burillo:2014aa}.
In the cases of B or C (Figure~\ref{fig:schematic}), the motion of molecular gas of NGC~1316 can be (partially) explained by outflow.
As previous studies with dust data claimed \citep{Geldzahler:1984aa,Lanz:2010aa}, the nuclear jet seems to bend at just south of the NW Shell and just north of the SE Blob (Figure~\ref{fig:comparison}f), implying a interaction between ISM and nuclear jet (see Section~\ref{sec:bending}).
In addition, we showed that the observed GDR and H$_2$-to-\HI~ratio of NGC~1316 are comparable to those of galaxies with stellar mass of $>10^{11.5}$~M$_\odot$, which is comparable to the stellar mass of NGC~1316.
The molecular gas fraction, $f_{\rm mol} = \frac{M_{\rm mol}}{M_{\rm mol}+M_\star}$ (where $M_\star$ is stellar mass) of NGC~1316 is $\sim 0.6$~\%, which is not so high compared to those of ETGs with similar stellar masses \citep{Young:2014aa}.
The smaller velocity gradient (outflow velocity) of molecular gas in NGC~1316 ($\sim |200|$~km~s$^{-1}$) compared to the typical galaxies with molecular outflow ($>|500|$~km s$^{-1}$) cannot be a strong objection against the internal-origin scenario, because it is possible that the nuclear jet has a small inclination ($\lesssim 24$ degree).
The small inclination of the nuclear jet is suggested from the symmetric structure and the intensity ratio between the north and in the south.
Therefore, these properties may prefer outflow scenario for molecular gas in NGC~1316.

However, there are also some facts against the outflow scenario for NGC~1316.
First, the simplest speculations on the relative line-of-sight locations of the NW Shell, the SE Blob and the Nucleus from the dust-extinction and the CO or dust emission data is the case A, suggesting inflow motion, as shown in Figure~\ref{fig:schematic}.
Second, the dust-to-stellar mass ratio of NGC~1316 is unusually high (i.e., dust rich) compared to the other early-type galaxies with similar stellar masses \citep{Lanz:2010aa}.
Finally, the steep velocity gradient along the NW Shell (PV cut ``A'' in Figure~\ref{fig:pv_maps}) is difficult to reproduce in the jet-induced molecular outflow.

In reality, the situation is not so simple and all the factors discussed in sections~\ref{sec:external} and \ref{sec:internal} may play a role to make the ISM in the current physical status.
The gas may be injected by other galaxies and blown out by AGN jet, or the internal origin gas is first blown out by AGN jet then flows back to the nucleus.
In addition, it should be noted that the flux ratio of CO to Herschel/PACS~$160\mu$m data ($\sim$~GDR) at the NW Shell is roughly two times higher than that of the SE Blob.
This may suggest that the degree of jet-ISM interaction is different for the NW Shell and the SE Blob or that it is not a single galaxy which injected the molecular gas and dust.
Again, with these observational boundary conditions, it is important to investigate the detailed history of galaxy merger of NGC~1316 with numerical simulations of galaxies with molecular gas and dust formation and AGN feedbacks.

\section{Bending of the nuclear jet}
\label{sec:bending}

Jet bending has been observed in some radio galaxies \citep[e.g.,][]{Ekers:1978aa,Smith:1981aa,Wilson:1982aa}.
Figure~\ref{fig:comparison}(f) showed that the nuclear jet bends at the vicinity where the dust and molecular gas emission are strong, regardless of the origin of the molecular gas of NGC~1316.
Three scenarios have been claimed to explain the bending jets:
(1) precession of the jet nozzle \citep[e.g.,][]{Ekers:1978aa},
(2) buoyant effect \citep[pressure gradient of the ISM, e.g.,][]{Smith:1981aa,Henriksen:1981aa,Bridle:1981aa} or
(3) ram pressure effect from ISM \citep[e.g.,][]{Wilson:1982aa}.
\cite{Geldzahler:1984aa} considers that (1) and (3) are unlikely:
for scenario (1), the symmetry of the nuclear jet is not correct for the precession and the orientation of the nuclear jet appears to be associated with kpc-scaled phenomenon \citep[e.g., the X-ray cavity][]{Lanz:2010aa};
for scenario (3), the direction of the rotation of the gas disk implied from the ionized gas and the dust extinction data is opposite to the one expected in the case of ram pressure.



Our data clearly showed that the jet bends at the denser regions (higher pressure), toward sparser regions (lower pressure) of molecular gas, which supports the (2) buoyant effect scenario.
Whether or not the (3) ram pressure scenario is ruled out depends on the kinematics of molecular gas.
In all cases (A, B, and C), the direction of the jet bending is consistent with the flow direction of the molecular gas (from east to west for the NW Shell and vice versa for the SE Blob).
Thus, our ALMA data support that both scenarios (2) and (3).

\section{Summary}
\label{sec:summary}

We conducted a $^{12}$CO($J$=1-0) mapping observation of NGC~1316 as part of the ALMA survey of 64 Fornax galaxies and also optical spectroscopic observation with Keck/LRIS.
The obtained results and implications based on the data are as follows:
\begin{itemize}
\item The obtained total molecular gas mass ($5.6 \times 10^{8}$~M$_\odot$ with the standard Milky-Way $X_{\rm CO}$) and the velocity gradient between the NW Shell and SE Blob ($\sim 400$~km~s$^{-1}$) are consistent with previous single-dish observations within the error (section~\ref{sec:results} and Figure~\ref{fig:pv_maps}).

\item $^{12}$CO($J$=1-0) distribution coincides with dust structures seen in the previous high-resolution optical and FIR observations (Figure~\ref{fig:comparison}).
For the first time, our high spatial resolution and deep CO data reveal the NW shell structure, some clumps between the NW Shell and the SE Blob, and the SE extended structure just above the Blob (section~\ref{sec:distribution}).

\item We found some features disfavoring a simple rotating disk scenario for the central cold ISM of NGC~1316, such as the crossing of the systemic velocity of NGC~1316 in the NW Shell, the multiple velocity components in the NW Shell (PV cut ``A'' in Figure~\ref{fig:pv_maps}), the monotonic velocity variation in the NW Shell and the SE Blob (PV cuts ``A'' and ``F''), and the velocity deviations from the interpolation between the velocities of the NW Shell and the SE Blob (PV cut ``C'', section~\ref{sec:velocityfield}).

\item Keck/LRIS observation confirms that H$\alpha$ emission is very weak in NGC~1316, as the previous optical study indicates.

\item Despite some disagreements in kinematics of molecular and ionized gas, the velocity of ionized gas estimated with \oii~line is roughly consistent with that of molecular gas: $\sim1500$~km~s$^{-1}$
at 25 arcsec in SE and $\sim1900$~km~s$^{-1}$ at 50 arcsec in NW from the galaxy center.
Although the accuracy of determining the central velocity of ionized gas is not so high ($\sim100$~km~s$^{-1}$), the obtained PV map of \oii~cannot be explained only by rotation.

\item 
If the observed molecular gas has an external origin, the complex spatial distribution and velocity field suggest a recent merger of $\lesssim 1$~Gyr.
The inferred stellar mass of the merged galaxy from the observed H$_2$-to-\HI~($\sim5.6$) and dust-to-gas ratios (GDR$\sim28$) is $>10^{11.5}$~M$_\odot$, while it is unlikely that NGC~1316 is {\it now} experiencing a major-major considering the ordered distribution of its stellar component.
To explain the observed H$_2$-to-\HI~ratio and GDR, additional processes should be taken into account such as an effective dust formation and conversion from atomic to molecular gas during the interaction (section~\ref{sec:internal}).

\item The nuclear jet bends at the NW Shell and the SE Blob suggesting an interaction between the jet and ISM.
Our data support the scenario that the nuclear jet is bent due to the buoyant effect and/or ram pressure from the background ISM (section~\ref{sec:bending}).

\end{itemize}





\begin{ack}
We thank the anonymous referee for his/her comments which improve our paper.
We are grateful to Dr.~Nao Suzuki and Dr.~Saul Perlmutter for kindly providing us with Keck/LRIS data of NGC~1316.
KMM acknowledges Dr.~Asao Habe and Dr.~Hiroshi Nagai for meaningful discussions on the evolution of gas in the elliptical galaxies and the interpretation of the observed kinematics of molecular gas in NGC~1316, respectively.
This paper makes use of the following ALMA data: ADS/JAO.ALMA\#2017.1.00129.S.
ALMA is a partnership of ESO (representing its member states), NSF (USA) and NINS (Japan), 
together with NRC (Canada), MOST and ASIAA (Taiwan), and KASI (Republic of Korea), in 
cooperation with the Republic of Chile. The Joint ALMA Observatory is operated by 
ESO, AUI/NRAO and NAOJ.
This research has made use of the NASA/IPAC Extragalactic Database (NED),
which is operated by the Jet Propulsion Laboratory, California Institute of Technology,
under contract with the National Aeronautics and Space Administration.
This publication makes use of data products from the Wide-field Infrared Survey Explorer, which is a joint project of the University of California, Los Angeles, and the Jet Propulsion Laboratory/California Institute of Technology, funded by the National Aeronautics and Space Administration.
This project has received funding from the European Research Council under the European Union's Horizon 2020 research and innovation programme (grant agreement no. 679627; project name Fornax).
Parts of this research were conducted with the support of Australian Research Council Centre of Excellence for All Sky Astrophysics in 3 Dimensions (ASTRO 3D), through project number CE170100013.
T.M. was supported by JSPS KAKENHI Grant Number JP 16H02158.
D.E. was supported by JSPS KAKENHI Grant Number JP 17K14254.
F.E. was supported by JSPS KAKENHI Grant Number JP 17K14259.
\end{ack}


\bibliographystyle{apj}
\bibliography{/Users/kanamoro/Documents/BibFile/myref_moro}

\end{document}